# iAQC: The Intensity-Aware Quantum Cryptography Protocol

Subhash Kak, Yuhua Chen, Pramode Verma

**Abstract:** This paper reports a variant of the three-stage quantum cryptography protocol which can be used in low intensity laser output regimes. The variant, which tracks the intensity of the laser beam at the intermediate stages, makes the task of the eavesdropper harder than the standard K06 protocol. The constraints on the iAQC protocol are much less than those on BB84 and in principle it can not only be used for key distribution but also for direct bitwise encryption of data. The iAQC protocol is an improvement on the K06 protocol in that it makes it harder for the eavesdropper to monitor the channel.

## Introduction

We present here a variant of the three-stage quantum cryptography protocol [1], sometimes called the K06 protocol, which has protection against eavesdropping built into it. The 3-stage quantum cryptography protocol is based on random rotations (or other commutative operators) which can better protect duplicate copies of the photons than in non-single qubit transmissions of BB84. This protocol can use attenuated pulse lasers rather than single-photon sources in the quantum key exchange, which makes it possible to transmit the pulses over a greater distance. The theoretical basis of the K06 protocol is the fact that unknown pure states carry information [2],[3],[4] even though the von Neumann entropy of such states equals zero.

In this new protocol, which we call iAQC protocol, both Alice and Bob monitor the intensity of the light beam coming to them in the intermediate stages of the protocol. Doing so makes it possible for Alice and Bob to determine if any photons have been siphoned off by the eavesdropper. This tracking can be done by measuring a pre-set fraction of the incoming beam. This makes the task of Eve, the eavesdropper, harder than it would be without the intensity awareness feature. If intensity is not tracked then iAQC is identical to K06.

The development of this protocol is motivated by search for effective implementations of the K06 protocol, which is a collaborative project of Oklahoma State University, University of Oklahoma Tulsa, and University of Houston that is funded by the National Science Foundation. A major attribute of the three-stage protocol is that the unitary transformations used by Alice and Bob can change as fast as one cycle time that a photon takes in traversing the round-trip distance from Alice to Bob and back. This means that Alice and Bob can independently and concurrently change their transforms, making it all but impossible for an intruder to siphon off photons in transit and making any sense out of them. The basic free-space implementation of the K06 protocol has been achieved and current research is on increasing data rates and implementation in fiber optics.

This paper will first provide background material on the K06 protocol and then describe the iAQC protocol.

## K06 Protocol

The K06 protocol is described as follows. Consider transferring qubit state X from Alice to Bob. The state X is one of two orthogonal states and it may represent 0 and 1 by prior agreement of the parties, and this is the quantum cryptographic key being transmitted over the public channel (X can also be a classical bit, that is 0 or 1). Alice and Bob apply secret transformations $U_A$ and $U_B$ that are commutative. These secret transformations can be changed as frequently as desired.

> *Step 1:* Alice applies a unitary transformation $U_A$ on quantum information X and sends the qubits to Bob.
>
> *Step 2:* Bob applies $U_B$ on the received qubits $U_A(X)$, which gives $U_B U_A(X)$, and sends it back to Alice.
>
> *Step 3:* Alice applies $U_A^\dagger$ (transpose of the complex conjugate of $U_A$) on the received qubits to get $U_A^\dagger U_B U_A(X) = U_A^\dagger U_A U_B(X) = U_B(X)$ (since $U_A$ and $U_B$ commute) and sends it back to Bob.
>
> *Step 4:* Bob applies $U_B^\dagger$ on $U_B(X)$ to obtain X.

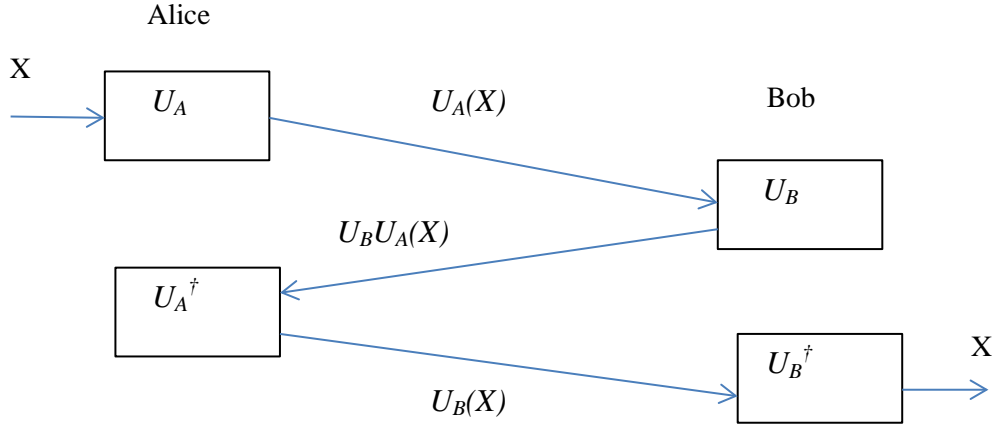

**Figure 1.** The K06 protocol (the transformations are rotations)

The use of random transformations, which Alice and Bob can change from one qubit to another, guarantees that from the perspective of the eavesdropper, the probability of collapsing into $|0\rangle$ and $|1\rangle$ states has equal probability, which is desirable for cryptographic security.

As we can see, while the actual quantum state of X is never exposed on the link, Bob is able to restore X. One can use the commutativity of the rotation operator $(\theta) = \begin{pmatrix} \cos\theta & -\sin\theta \\ \sin\theta & \cos\theta \end{pmatrix}$, clear from the relation:

$$R(\theta) \cdot R(\phi) = \begin{pmatrix} \cos\theta & -\sin\theta \\ \sin\theta & \cos\theta \end{pmatrix} \cdot \begin{pmatrix} \cos\phi & -\sin\phi \\ \sin\phi & \cos\phi \end{pmatrix} = \begin{pmatrix} \cos(\theta+\phi) & -\sin(\theta+\phi) \\ \sin(\theta+\phi) & \cos(\theta+\phi) \end{pmatrix}$$

for implementing $U_A(X)$ and $U_B(X)$ as $R(\theta)$ and $R(\varphi)$.

Unlike the BB84 protocol which is vulnerable to siphoning of photons in an attenuated pulsed laser system, the 3-stage protocol is immune to such an attack since the actual quantum state of the key is never revealed in the communication. This property is of significant importance in terms of using quantum cryptography in a practical network environment where an optical path can potentially be extended beyond trusted routers.

## iAQC Protocol

In iAQC, the intensity-aware protocol, the intensity of the beam generated by Alice is publicly announced [5]. For the sake of simplicity of presentation it will be assumed that there is no loss in intensity during transmission from Alice to Bob. In practice, this loss can be factored in the protocol. The protocol is as follows:

*Step 1*. Alice uses photons of intensity I (which is publicly known) to Bob and performs a polarization rotation of angle θ.

*Step 2*. Bob uses a partially silvered mirror (or some suitable beam-splitter) to divert fraction *k* of the incoming beam to measure its intensity. He applies another random rotation φ on the remaining beam of intensity *I(1-k)* and transmits it to Alice.

*Step 3*. Alice receives the beam of intensity *I(1-k)* and diverts a fraction *k* of it to measure its expected intensity and undoes her random rotation of angle θ and retransmits the beam of intensity *I(1-k)²* to Bob.

*Step 4*. Bob uses the intensity-testing portion of it *I(1-k)²k* for checking it and uses the rest of the beam of intensity *I(1-k)³* to determine the value of the qubit.

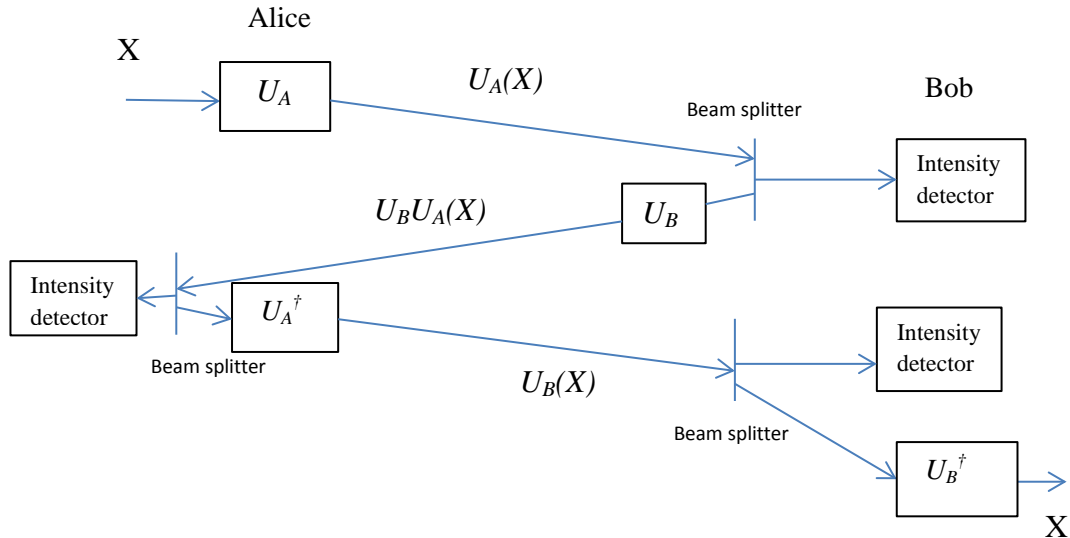

**Figure 2.** iAQC Protocol

The eavesdropper will be detected by Alice and Bob if the intensity of the beam reaching them in the three stages is less than expected.

The constraint on the maximum number of photons being used in the transmissions between Alice and Bob is much less severe than in BB84. If a single photon is transmitted then the system is provably secure. Let us assume that Alice and Bob each uses $s$ different rotation angles. Eve needs to have access to each of the three transmissions to be able to break the system. Assuming that she needs $m$ photons to determine which one of the $s$ rotation angles were used by Alice and Bob, the total number she needs to siphon off is $3m$. If the photon source produces $6m$ photons, the siphoning off of $3m$ photons would reduce the intensity by a factor of 2. The intensity used could be a clearly defined variable of the system and any reduction in it would be obvious to Bob and therefore the eavesdropper would be found out.

Also note that $m \gg s$, therefore, the laser source output can consist of many photons. From information theoretic point of view, the number of photons required to distinguish between $s$ rotation angles is $\log_2 s$. Therefore, the minimum number of photons that need to be siphoned off by the eavesdropper is $3\log_2 s$. In reality, the number of photons to be siphoned off will be significantly greater.

If the technique of using s photons each fed into detectors aligned to different polarizations is used then one requires a total of $3s$ photons to be siphoned off by Eve to determine the three different polarization angles. If the detectors used by Bob and Alice to check the intensities at intermediate stages can determine reduction of intensity by a factor of one-half, then the source can use $6s$ photons in each burst. In reality the constraints are even less severe as the bases used by Alice and Bob are arbitrary as long as they are in the same plane.

## Siphoning Attack on K06 and iAQC

Eve can fool Alice and Bob into believing that no photons have been siphoned off if she injects the number of random photons equal to the ones she has taken from the beam. But this can only be done at a cost to her ability to tap the exchange. If she withdraws g percent of the photons and injects the same number of some known polarization, the probability that she will be able to determine the polarization angles of the siphoned photons will go down in the next two passes. Furthermore, Bob would know that siphoning had taken place from the fact that the measured photons will not all have the same polarization.

To see this, assume that Alice sends 6 photons and Eve takes out 1 photon in each link. The following example picture describes the situation:

Table 1. The workings of the iAQC protocol

| Alice Sends | A(X) | A(X) | A(X) | A(X) | A(X) | A(X) |
|---|---|---|---|---|---|---|
| Photons after Eve's first siphoning | A(X) | A(X) | A(X) | A(X) | A(X) | E |
| Bob Sends Back | BA(X) | BA(X) | BA(X) | BA(X) | BA(X) | B(E) |
| Photons after Eve's second siphoning | BA(X) | BA(X) | BA(X) | E | BA(X) | B(E) |
| Alice Sends to Bob | B(X) | B(X) | B(X) | $A^{-1}(E)$ | B(X) | $A^{-1}B(E)$ |
| Photons after Eve's third siphoning | B(X) | B(X) | B(X) | $A^{-1}(E)$ | E | $A^{-1}B(E)$ |
| Photons obtained by Bob | X | X | X | $B^{-1}A^{-1}(E)$ | $B^{-1}(E)$ | $A^{-1}(E)$ |

Bob would be easily able to determine that Eve had siphoned off the photons since all his photons will not be aligned.

## Conclusions

The BB84 protocol for quantum key distribution (QKD) and its variants [6],[7] deal with the secure exchange of key between sender and utilizing the quantum properties of photons. The major threat to the security of QKD algorithms comes from the fact that the constraints on the implementation of the optical apparatus used in these protocols are extreme. BB84 is provable secure only if the photon sources produces single photons [8] and the detector can detect single photons. Recently, industry implementations of the protocol that were thought to be secure were successfully hacked [9],[10]. Although patches for these attacks have been introduced, there is no evidence at this point that other loopholes in the implementation do not accept.

In general, BB84 remains open to siphoning attacks in some form or the other if the number of photons being transmitted per time unit exceeds one. On the other hand, if the average photon rate is much less than one as used in the implementation schemes of idQuantique and MagiQ, the overall rate of a few thousand bits per second makes the system unsuitable for quantum cryptography of continuous data and that is the reason why it is used only for quantum key distribution. Current systems employ symmetric classical cryptography after the key has been exchanged using BB84.

The constraints on the iAQC protocol are much less than those on BB84. The iAQC protocol is an improvement on the K06 protocol in that it makes it harder for the eavesdropper to monitor the channel. An analysis of the performance of the iAQC protocol in the presence of noise remains to be done. For this analysis further information on the number of photons used and capability to determine polarization states will be required [11].

**Acknowledgements.** This research is supported in part by the National Science Foundation (NSF) under Grants 1117148, 1117179, and 1117068.

**REFERENCES**

[1] S. Kak, A three-stage quantum cryptography protocol. Foundations of Physics Letters 19, 293-296 (2006).

[2] S. Kak, Quantum information and entropy. International Journal of Theoretical Physics 46, 860-876 (2007).

[3] S. Kak, The transactional nature of quantum information. 11th International Conference on Squeezed States and Uncertainty Relations and 4th Feynman Festival, Olomouc, Czech Republic, June 22-26, 2009.

[4] S. Kak, P. Verma, and G. MacDonald, Cryptography and state estimation using polarization states. SPIE Conference on The Nature of Light: What are Photons IV? August 2011.

[5] M. Lavale, Security strengthening of the three-stage quantum key distribution protocol. M.S. Thesis, Oklahoma State University, May 2012.

[6] C. H. Bennett and G. Brassard, Quantum cryptography: Public key distribution and coin tossing. Proceeding of the IEEE International Conference on Computers, Systems, and Signal Processing, Bangalore, India, pp. 175–179 (IEEE, New York, 1984).

[7] C. H. Bennett, Quantum cryptography using any two nonorthogonal states. Physical Review Letters 68, 3121-3124 (1992)

[8] T. Usuki, Y. Sakuma, S. Hirose, K. Takemoto, N. Yokoyama, T. Miyazawa, M. Takatsu, Y. Arakawa, Single-photon generator for optical telecommunication wavelength. Journal of Physics: Conference series, vol. 38 140, 2006.

[9] I. Gerhardt, Liu, Q., Lamas-Linares, A., Skaar, J., Kurtsiefer, C., and Makarov, V. Full-field implementation of a perfect eavesdropper on a quantum cryptography system. Nat. Commun. 2, 349 (2011)

[10] L. Lydersen, Wiechers, C., Wittman, C., Elser, D., Skaar, J. and Makarov, V. Hacking commercial quantum cryptography systems by tailored bright illumination. Nat. Photonics 4, 686 (2010)

[11] D.S. Talaga, Information theoretical approach to single-molecule experimental design and interpretation. J Phys Chem A. 110(31): 9743–9757 (2006)